# Heterarchy in the Brain: Control as a Spectrum, Not a Chain of Command

Luiz Pessoa[1] and Andrea Gambarotto[2]

(1) Department of Psychology and Maryland Neuroimaging Center, University of Maryland, College Park, USA

(2) Department of Humanities, Université du Luxembourg, Luxembourg

## Highlights

- Brain-wide cellular recordings and graded analyses increasingly reveal functional populations that cross cytoarchitectural boundaries.
- Causal perturbation and directed-connectivity studies show that reciprocal circuits can reverse their dominant direction across task phases and states.
- Research on parallel action systems and affordance-sensitive behavior shows that control can migrate while displaced solutions remain available.
- Dynamical and organism-centered theories increasingly recast control as context-sensitive coordination under urgency and viability demands, rather than executive command.

## Abstract

Neuroscience treats the brain as a control hierarchy, with cortical systems governing subcortical mechanisms. We argue that this mistakes a configuration for an architecture. Local asymmetries of influence are real, but control is relational and process-specific; because neural elements constrain one another across concurrent processes, relations can form cycles that resist a stable ranking. Evidence from learned action, cortico-cerebellar and thalamocortical loops, and defensive behavior shows that control changes with practice, task phase, urgency, viability demands, and affordances. Hierarchical order therefore characterizes time-bounded configurations, whereas heterarchy characterizes the architecture's repertoire across occasions. We propose a spectrum of control indexed by dynamical decoupling: how far behavior can depend on history, internal state, and anticipated consequences. The account yields experimentally testable predictions.

# 1. From the network model of emotion to the problem of control

A decade ago one of us argued that emotion is not housed in a circumscribed set of regions but is a property of **functionally integrated systems** spanning the neuroaxis [1,2]. That argument attempted to dissolve one boundary: between “the emotional brain” and the rest of the brain. But where a function lives and what governs it are separate questions. One can grant that emotion is distributed across cortex, amygdala, hypothalamus, and midbrain and still hold that cortex governs it, and most of the field does. This Opinion challenges the idea of *hierarchical control*, a notion that reaches well beyond emotion.

A hierarchical picture of control is easy to dismiss as caricature: few now defend a cortex dictating to a passive, “reptilian” subcortex. Yet a consequential version persists—the assumption that structures producing a behavior occupy a stable control order, with cortex constraining systems beneath it and feedback modulating rather than overturning that order. This assumption extends beyond emotion to action, motivation, and cognitive control. It appears in canonical accounts of prefrontal cortex as the source of top-down control [3,4], emotion regulation as prefrontal control of the amygdala and psychopathology as failure of that control [5–7], and predictive-coding hierarchies in which higher cortical levels constrain lower ones [8,9]. Although many have challenged the picture [10–14], it remains a default that shapes interpretations of connectivity and lesions, as well as clinical narratives.

In cognitive science, control ordinarily means **cognitive control** (see Glossary): the maintenance of goals, the resolution of conflict, the suppression of prepotent responses, capacities conventionally attributed to prefrontal cortex. The sense at issue here is broader. In the theory of biological autonomy, a control mechanism is one that does not carry out a process itself but determines when and how other mechanisms operate, and does so in the service of the organism’s own conditions of persistence [15–17]. Bacterial chemotaxis is control in this sense [17] but does not involve what one would call cognition. We call this **biological control**. Where we write control without qualification below we mean the biological sense; cognitive control is named in full.

What control looks like in a brain is easiest to see by asking what problems it must solve. Animals face problems spanning enormous urgency, from a predator at contact, demanding action within tens of milliseconds, to one somewhere in the landscape, where the problem is how to forage for an hour without meeting it. Ethology conceptualizes defensive behavior along this continuum of predatory imminence [18,19]. It is tempting to reserve control, goals, and decisions for the leisurely end. Yet an animal fleeing at quasi-contact still selects among routes according to what the terrain affords [20]; integration of signals, evaluation of costs, and resolution of conflict remain present under time pressure and tightly tied to bodily state [19,21]. The difference along the range is one of temporal scale and urgency.

The hierarchical picture rests on a premise that has rarely been stated precisely: among the structures supporting a given behavioral capacity, the local relations compose into a consistent ordering, and that ordering belongs to the system rather than to the occasion, so context modulates the strength of the relations without reversing their direction. Almost no one holds the stronger claim, that every structure in the brain can be ranked against every other. At the same time, narrower and far more widely held views license reading function off anatomy, naming the controller of a behavior, and grounding the notion that cognition governs emotion.

This Opinion defends the thesis that control in the brain is *heterarchical* rather than hierarchical: local, directional asymmetries of influence are real and systematic, but they do not assemble into stable orderings. Two factors settle which structure governs: the animal's history determines which control relations are available to it, and the occasion selects among those. Hierarchy, where it holds at all, is the special case and heterarchy the general condition. Rejecting a standing controller creates a selection problem: what makes one configuration rather than another govern now? Circuits differ in how far a response can depend on more than the situation immediately present, and that degree of **dynamical decoupling** is the spectrum of control: at one end a response follows the situation in front of the animal, at the other it waits on what the animal remembers, wants, and has learned. Urgency, weighed against the organism's **viability norms** [22,23], sets where an occasion falls on the spectrum.

One clarification is in order. That the brain has levels of organization—molecules, cells, circuits—is not in dispute, nor is the local asymmetry between a **constraint** (see Glossary) and what it regulates. A constraint channels a process while remaining unchanged on that process's timescale [24]: riverbanks shape a river; connectivity fixes which structures can directly influence which; synaptic weights and modulatory state determine what propagates and how strongly. None carries the signal; all shape it. A mechanism that acts on other constraints rather than directly on flows of matter and energy—a second-order constraint, the autonomy-theoretic sense of regulation [15]—is what separates control from **dynamic stability**. But the relation is local. A dimmer emits no light, so for the lamp's illumination it is second-order; for the circuit, it is one more resistance in the line. A mechanism's order therefore depends on the process followed and cannot by itself rank one structure above another. A **control hierarchy**, by contrast, would rank the structures supporting a behavioral capacity by command, stably across occasions—for example, a cortical "cognitive" process governing a noncortical "emotional" mechanism. "Global" below means across those structures and occasions, not the entire brain. For historical background, see Box 1.

## 2. Why local asymmetries need not compose

Begin with the intuition that makes hierarchy attractive. If mechanism A regulates a process carried out by B, we place A "above" B, and stacking such relations seems to yield a command structure. In a simple device that inference is sound: the lamp never sets the terms on which the dimmer operates, so controller and controlled occupy fixed positions (Fig. 1A). Reading the brain in the same way—with cortex as the sophisticated switch and the rest as the lamp—is tempting. The step, however, is invalid.

The brain does not work like that because a single element can occupy different relational positions at once, at multiple measurable levels. Paraventricular thalamic neurons send collaterals to nucleus accumbens, bed nucleus of the stria terminalis, and central amygdala, so one cell belongs to several circuits before activity is measured [25]. Recordings of tens of thousands of mouse neurons during decision-making find that each region contains cells spanning nearly the full repertoire of task-related responses, with functional identity poorly predicted by cytoarchitectural boundaries [26], as anticipated by nonlinear mixed selectivity [27]; human imaging recovers overlapping circuit membership at coarser scale [28,29]. Participation is graded rather than all-or-none: the activity reaching a downstream area occupies a subspace distinct from that area's dominant local fluctuations, so the same neurons participate simultaneously in a local regime and an interareal channel, in measurable proportion [30].

None of this says the brain is equipotential. Regions differ in their connections and contributions, and the asymmetries catalogued in Section 3 depend on those differences. But a structure may set the terms for other processes while having its own terms set in turn: it is controller or controlled depending on the process. The same is true in any densely cross-linked network, where a junction is upstream of one route and downstream of another. Thus "above" and "below" are directions of travel, not fixed addresses. This alone is insufficient for heterarchy: the relations must be relations of control and must fail to compose within the behavioral capacity and occasions being explained.

Could local orderings still compose into a hierarchy? Yes, but only if they fit a consistent partial order, which may include parallel branches while remaining *transitive* (if A outranks B and B outranks C, then A outranks C) and free of loops. A single loop defeats it: if A sets the terms for B while B sets the terms for A, no assignment of rungs satisfies both.

The argument is: (i) a structure is a controller only relative to a process unfolding over a particular interval (assembling a response, holding two structures in phase, or changing a synapse); (ii) each element participates in multiple processes, regulating some while regulated in others; and (iii) these relations form loops. Because the processes share elements, an ordering established for one cannot be assumed for another. Constraint order is real but local; it does not compose into a control hierarchy (Figure 1B).

Premise (iii) is critical and must not be confused with reciprocal wiring. Hierarchies with feedback are coherent; a thermostat has feedback, so pathways running in both directions demonstrate nothing by themselves. The premise requires structures to constrain one another's processes: the amygdala shapes what prefrontal cortex has to work with while prefrontal cortex shapes what the amygdala works on; the periaqueductal gray instructs plasticity in structures that also instruct it. The claim is not that everything influences everything equally, but that these directional asymmetries do not compose. Bidirectional influence and convergence make such cycles an expected empirical pattern once reciprocal constraint is demonstrated, undermining a linear "billiard-ball" account in favor of mutual determination [2]. McCulloch anticipated this failure of local orderings to assemble globally (Box 1).

One more ingredient is needed before "control" does real work, because not every mechanism that keeps a process on track controls it. A mechanism that reacts only to the present state of what it acts on, pushing back as soon as that state drifts, merely holds the process steady, however elaborate its feedback (dynamic stability). Control is different. A controlling mechanism has dynamics of its own, so it can respond differently to the same current state depending on the wider situation, the organism's internal state, and what it has learned [15,17]. An idealized reflex has no such independence: the same stimulus produces the same response whatever else is true of the animal. Real reflexes have some: fear potentiates startle, so the same sound produces a larger response when a shock-associated cue is present [31,32]. This relative independence, or decoupling, comes in degrees and provides the main axis of the proposal below. One plausible evolutionary interpretation is that vertebrate brains increased this capacity by interposing additional circuitry between sensation and action [2,12,33–35].

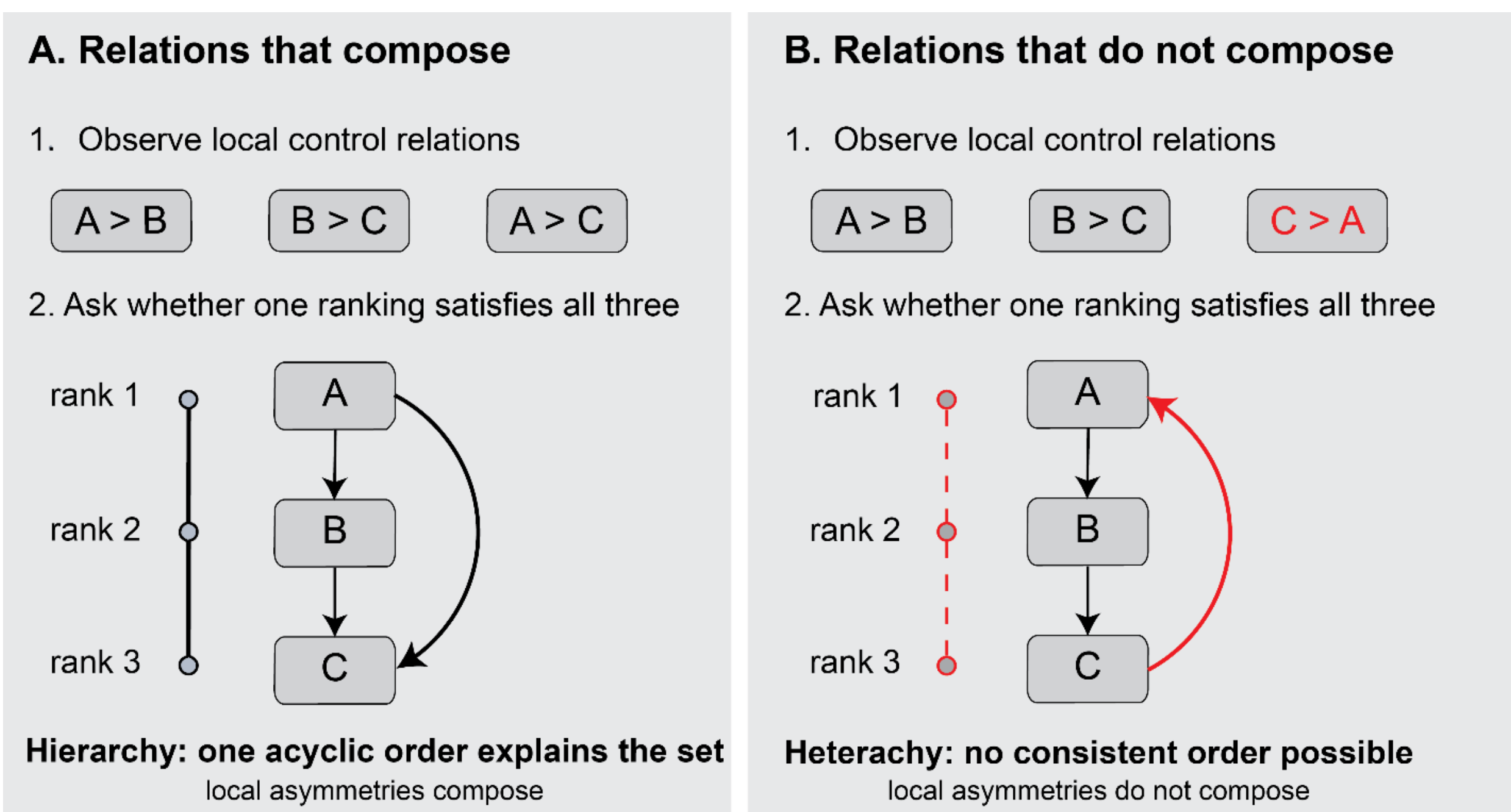

**Figure 1. Local control asymmetries need not compose into a hierarchy.** (A) A control hierarchy: influence relations form a directed acyclic graph, so every structure can be assigned a consistent level (a "chain of command"). (B) A heterarchy: the same structures constrain one another's processes, producing cycles ($A \rhd B$, $B \rhd A$; and the three-cycle $A \rhd B \rhd C \rhd A$) that admit no consistent level assignment.

## 3. Control changes hands across behavioral domains

A useful test must do more than identify two circuits supporting the same behavior, because a hierarchy can contain parallel branches. It must show that the operative ordering belongs to the occasion rather than the anatomy: for example, a displaced solution remains available, causal direction changes with task phase, or a structure cast as subordinate sets the terms under which its supposed superior operates. The examples that follow span different domains and timescales, asking in each case whether control is fixed by anatomy or reorganized by task demands, learning, and context.

**Practice shifts control**

During training, rats started from the same position and found food in the same goal arm, so reaching it always required the same turn at the junction [36]. This made two strategies indistinguishable: return to the same place or repeat the same turn. On a probe, rats started from the opposite arm. Returning to the learned place now required the opposite turn, whereas repeating the trained turn led elsewhere. Early in training rats returned to the place; after extended training they repeated the turn. Inactivation confirmed the causal dissociation: hippocampal inactivation disrupted the early place strategy, whereas caudate inactivation disrupted the later turn strategy. After the shift, caudate inactivation returned rats to the place strategy, showing that the displaced solution remained available. Microdialysis during the same navigation task showed that hippocampal acetylcholine rose roughly 60% at training onset and remained elevated after behavior shifted to the turn strategy. Striatal acetylcholine rose later, coincident with that shift [37]. Thus, strategy expression changed without a decline in this marker of hippocampal engagement.

Other evidence undermines the claim that extended practice transfers an action from a goal-directed to a habitual controller. After extensive lever-sequence training, neurons in both the dorsomedial and dorsolateral striatum tracked the initiation, execution, and termination of the sequence [38]. This finding is relevant because the dorsomedial striatum is typically associated with goal-directed action, whereas the dorsolateral striatum is typically associated with habitual action. The continued involvement of the dorsomedial segment therefore challenges the idea that extended practice causes goal-directed circuitry to disengage completely. In humans, five outcome-devaluation experiments likewise found no increase in habitual control after overtraining [39]. Together with the reversible-maze result, these findings weaken the simple story in which one system yields as another takes over. Practice may instead alter the relative influence of goal-directed and habitual processes while leaving both available, rather than transferring permanent ownership of the action from one system to the other.

**Cortical loops have no fixed controller**

A cortical motor hierarchy traditionally places frontal cortex above the cerebellum: cortex prepares movement, whereas the cerebellum executes or corrects it. However, evidence from mice performing a delayed directional-lick task indicates a reciprocal relationship. Preparatory activity developed in both anterior lateral motor cortex and the fastigial cerebellar nucleus [40]. Silencing motor cortex abolished cerebellar preparation, showing that cortical activity helps initiate the cerebellar preparatory state. Conversely, perturbing the fastigial nucleus during the delay disrupted the later licking response without impairing movement execution and weakened cortical preparation (cerebellar outputs returned through the thalamus to the relevant cortical region). Thus, cortex initiated the preparatory state, but cerebellar feedback helped sustain it and shape the subsequent response. Neither structure occupied a fixed position in the control hierarchy.

Thalamocortical loops show related interdependence. In a spatial working-memory task, mice retained spatial information across a delay for a later choice [41]. Inhibiting the mediodorsal-thalamus-to-prefrontal pathway during the delay impaired memory maintenance, whereas inhibiting the reciprocal prefrontal-to-mediodorsal pathway impaired the subsequent choice. The direction of effective control therefore changed with task phase: thalamus supported cortical maintenance during the delay, while cortex influenced thalamic processing when the stored information had to guide behavior.

Attention provides two further examples. In mice cued to attend either vision or audition, the mediodorsal thalamus did not encode the attentional rule itself. Instead, it amplified local prefrontal interactions, allowing rule-specific cortical activity to persist [42]. In monkeys attending to one of several visual locations, the pulvinar synchronized connected visual cortical areas according to the current attentional focus, thereby regulating communication among them [43]. Thalamus therefore does not merely relay a cortical command; it sustains, amplifies, and coordinates the cortical activity on which performance depends [44]. Across these examples, control is reciprocal and state dependent: cortical circuits establish task-relevant representations, while cerebellar and thalamic circuits stabilize, coordinate, and selectively reshape those representations as behavior unfolds.

**Extinction is reciprocal, not imposed**

Safety learning is often framed as cortex regulating subcortically driven fear. Medial prefrontal lesions can impair conditioned-fear extinction [45], inviting the idea of a cortical brake on the amygdala. Yet extinction is new, context-dependent learning, not erasure: fear returns with changes in context, time, or renewed exposure to the aversive outcome [46]. Yet, the amygdala is not a passive target. Blocking its NMDA receptors dose-dependently impairs extinction learning [47]. More directly, basal-amygdala “extinction neurons” switch on during extinction training and are reciprocally connected with medial prefrontal cortex. Because the classic infralimbic cue response appears only at later recall [48], this amygdala population may function upstream of cortical consolidation [49]. Extinction is therefore formed and expressed through reciprocal amygdala–prefrontal interaction, not imposed by cortex on the amygdala.

Understanding of the role of the midline thalamus in extinction has advanced considerably. Prefrontal–hippocampal theta coherence tracks the suppression of conditioned fear. Inactivating nucleus reuniens reduces such coherence and brings extinguished fear back, whereas 8 Hz stimulation of reuniens limits later relapse [50]. Reuniens is therefore not simply positioned above prefrontal cortex or hippocampus; it is necessary for coherence between them. Its own activity also shifts with learning: early in extinction, when fear remains high, reuniens is dominated by 3–6 Hz power that tracks freezing, and this gives way to faster theta as fear subsides (the same band in which prefrontal–hippocampal coherence accompanies suppression [50]). Which dynamical regime reuniens occupies thus depends on the animal's position in learning, not on a fixed role of the nucleus (Fig. 2A, 4). In humans, whole-brain analyses show independent networks operating in parallel at different timescales [51]. Safety learning thus emerges from interactions among cortex, thalamus, hippocampus, and amygdala, rather than cognition pressing downward on emotion.

**Response urgency reassigns control**

At high urgency, supposed output structures perform control operations. In the midbrain, dorsal periaqueductal gray determines whether to escape by applying a synaptic threshold to threat-related input from superior colliculus [52]. Superior colliculus also contributes causally to learned, abstract categorization [53]. Periaqueductal gray represents threat probability and prediction error [54,55] and provides a teaching signal to the amygdala (via the thalamus); inactivating it attenuates shock-evoked amygdala responses and impairs fear learning [56,57]. Influence also descends. Learned expectation suppresses the shock-evoked midbrain response, and a central-amygdala projection mediates that suppression [56,58,59]. The amygdala thus regulates a midbrain signal that, in turn, regulates amygdala plasticity (Fig. 2A, 1).

Which pathway governs also depends on what threat escape affords. Dorsal premammillary hypothalamus drives escape vigor through periaqueductal gray [60], but recruits an anteromedial thalamic pathway when escape can exploit the terrain, such as a rope or blocks [20]. The same thalamic nucleus contributes to learning the predator context [61]. Thus, a direct hypothalamus–periaqueductal gray route and a thalamus-mediated route are both available, and environmental affordances determine which control relation becomes operative (Fig. 2A, 2).

**Learning relocates control across the neuroaxis**

Repeated presentations of an expanding overhead, looming visual stimulus extinguish innate escape dependent on the superior colliculus and periaqueductal gray [62]. Responses in these areas then remain attenuated even when defensive behavior is restored by pairing looming with shock, while ventral hippocampal ensembles carry the updated response. The hippocampus therefore appears to encode the learned context or policy that permits escape behavior without reinstating the original midbrain pattern. Thus, defense returns, but the original midbrain solution does not simply reappear unchanged; it has been replaced or constrained by a newly learned configuration (Fig. 2A, 3).

The reverse also occurs. Visual cortex teaches suppression of looming-evoked escape through a projection to the ventrolateral geniculate nucleus, but lasting plasticity is stored in that thalamic target [63]. Cortex guides the change; a subcortical structure maintains it. More generally, control can also shift with time: retrieval of conditioned fear depends on a prelimbic-amygdala projection shortly after learning, but on paraventricular thalamus a day later [64]. Together with instrumental action, these results show that history can move control toward or away from cortex, while earlier solutions may remain active, become latent, or be durably altered. “Higher” and “lower” describe directions within a particular transaction, not enduring properties of the structures involved.

Across the three domains, the decisive evidence is relational. The plus-maze probe reveals a latent controller when the dominant one is blocked; reciprocal perturbations close the cortico-cerebellar loop; task phase reverses thalamo-prefrontal direction; affordances recruit a different escape route; and learned expectation changes an ascending teaching signal. Each result admits a hierarchical description of the particular interaction examined. What fails is their composition into one ordering that survives practice, phase, and context. Macroanatomical asymmetries remain real [65–69], but they underdetermine the operative control relation. In essence, heterarchy is the repertoire of these time-bounded configurations, not the claim that influence is equal or structure irrelevant (Figure 2B).

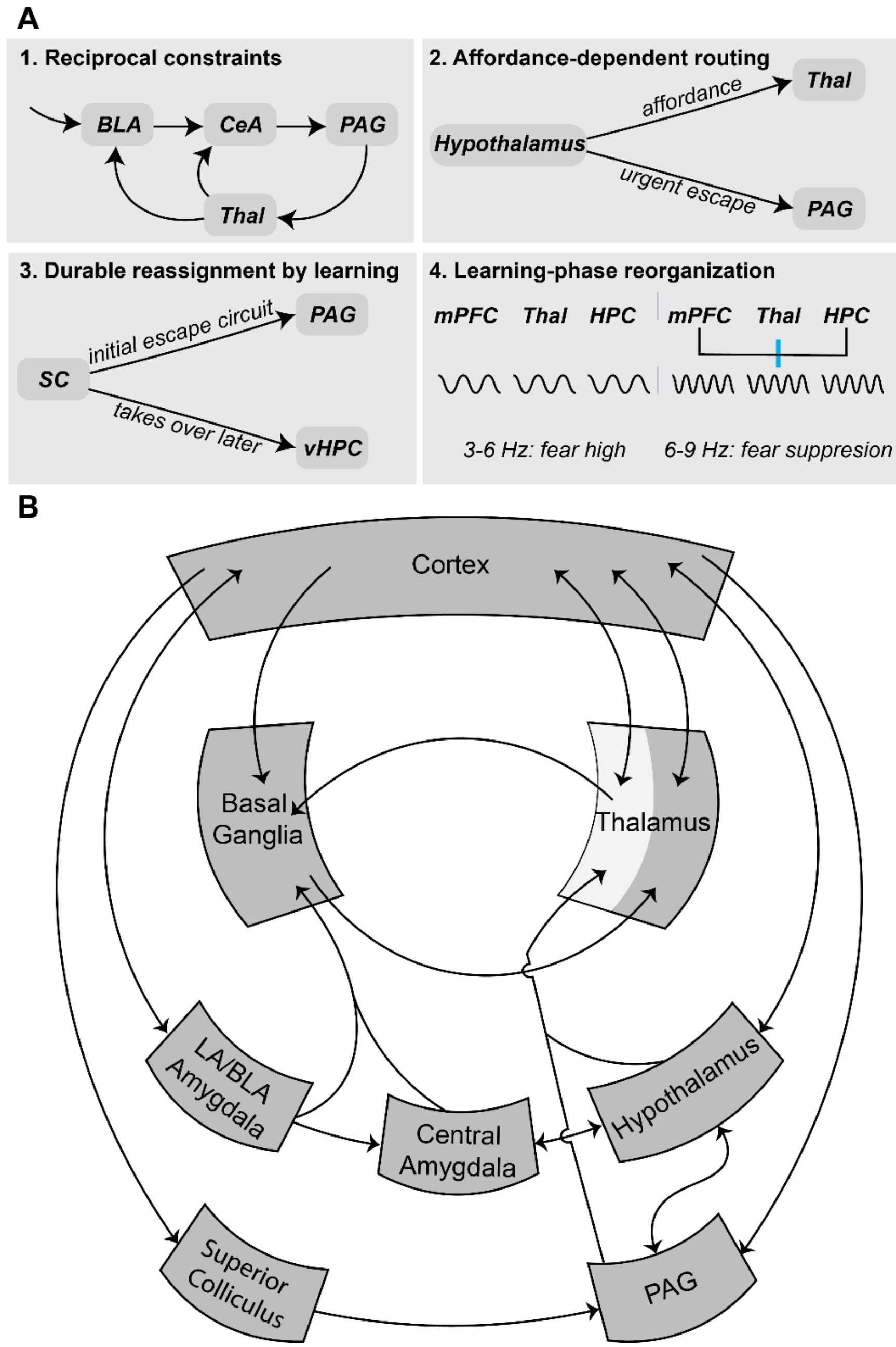

**Figure 2 (Key Figure). One architecture, many control configurations.**

**(A) Four ways the operative control relation changes while the anatomy stays fixed.** (1) Reciprocal constraints. Basolateral amygdala (BLA) drives central amygdala (CeA), which regulates periaqueductal gray (PAG); the PAG returns a teaching signal through thalamus (Thal) to BLA and CeA, so each structure shapes what the other learns or expresses [56,58]. (2) Affordance-dependent routing. Dorsal premammillary hypothalamus drives escape vigor through PAG, and recruits an anteromedial thalamic route only when the terrain offers something to exploit [20,60]. (3) Durable reassignment by learning. Escape initially depends on superior colliculus (SC) and PAG; after looming extinction those responses remain attenuated while a ventral hippocampal (vHPC) ensemble carries the restored response [62]. (4) Learning-phase reorganization. Early in extinction, with fear high, medial prefrontal cortex (mPFC), thalamus, and hippocampus (HPC) are dominated by 3–6 Hz activity; as fear subsides, 6–9 Hz coherence links the three, and nucleus reuniens (blue) coordinates that coupling [50].

**(B) The combinatorial, reciprocal, and reentrant neuroarchitecture that affords these configurations.** Cortex, basal ganglia, thalamus, amygdala, hypothalamus, superior colliculus, and PAG are joined by routes of differing length, many of them reciprocal.

# 4. The spectrum of control

Behavioral problems span a wide range of urgency, and one brain must handle them all. What varies is how far a response can depend on more than the situation immediately present. At the lower limit is the reflex, whose response is largely fixed by the stimulus. At the upper limit is behavior committed only after integrating remembered locations, physiological state, social rank, and anticipated consequences. The obvious way to cover this range is to build two systems, fast and slow, and let the slow supervise the fast. That is the hierarchical picture, requiring a stable ordering among the structures involved.

The mammalian macro-architecture offers a different arrangement, whose two salient features do different work (Figure 2B). It is combinatorial: many routes of different lengths lie between sensing and acting; and reciprocal and reentrant: structures along those routes constrain one another (the CRR architecture [70]). The combinatorial property affords the range—decoupling is purchased by interposing circuitry, not by adding a governor to an otherwise unchanged machine. Reciprocity and reentry leave the ordering open, because cyclic influence assigns no structure to a fixed level and the set engaged on one occasion can be reassembled on the next.

Heterarchy, on this view, is not an untidiness to be explained away. A fixed hierarchy could still produce multiple settings by varying the strength of its relations; flexibility alone therefore does not establish heterarchy. The narrower claim is that non-composition enlarges the repertoire by allowing the control relations themselves, not only their gain, to change across occasions.

Which setting obtains on a given occasion is not arbitrary (Box 2). It is constrained by the organism's norms of viability [22,23], with urgency as a central variable: decoupling buys flexibility but costs time, and time is what a threatened animal lacks. High urgency makes deferral expensive and biases commitment toward circuits with fewer integration steps and shorter paths to effectors; low urgency makes deferral viable, allowing more circuits to remain active in parallel before any one commits the system. Urgency in this sense is a relation between a situation and a self-maintaining organization rather than a property of the stimulus [15].

We hypothesize that no separate arbiter performs this selection. The proposal is close in spirit to accounts in which choice is settled by competition among the circuits that would carry out the alternatives, not by a dedicated comparator [71–74]. Circuits differ in how many integration steps lie between sensing and acting, and therefore in how long they take to commit. By commitment we mean entry into a state that causally determines the impending response. The circuit reaching such a state within the available time governs: a short deadline favors short-path, weakly decoupled circuits; a long deadline admits long-path, strongly decoupled ones. Constrain a response window to a few hundred milliseconds and participants express a practiced response they know to be wrong [75]. Path length is not the whole story: a circuit primed in advance can commit first, and evidence-accumulation rate matters as much as distance [76,77]. The claim is that available time biases which class of circuit commits.

A momentary configuration can nonetheless admit a correct hierarchical description for two reasons, neither about architecture. Influence along a diffuse, largely unreciprocated projection may be described locally as controller/controlled; along a strong, selective, bidirectional projection, it may not. And a configuration lasting tens or hundreds of milliseconds can exhibit definite asymmetries, making a control description adequate for that interval without supporting any claim about the system's standing organization. Both concern the scope of description, not what the brain is.

## 5. A testable signature

A network theory like the present one must answer Saper's objection: network descriptions are "a convenient explanation for complex responses," "tell us little about how they actually work," and tend "to stifle exploration for more parsimonious explanations" [78]. The objection has force and cannot be met rhetorically. This section addresses it.

If control is heterarchical as proposed, then among structures supporting a shared behavioral capacity: (a) dominant influence within a pair should reverse across contexts differing in urgency, learning history, or affordances; (b) each structure should shape what the other learns or expresses, beyond the bidirectional signaling guaranteed by reciprocal wiring; and (c) relations mapped across contexts should form a loop—A setting terms for B, B for C, C for A—so no ranking holds (Fig. 1B). A reliable reversal rejects stable pairwise ordering; a cycle across three or more relations establishes failure of composition. If instead the relations fall into a consistent ranking, with context changing only their strength, the hierarchical account is vindicated and this one is false. Section 3 assembles evidence from experiments designed for other purposes. Tests designed for this proposal are now needed.

Section 4 advanced a second claim: available time determines which circuit commits without requiring a separate arbiter. That claim carries a sharp test. Hold sensory evidence, learned contingencies, and response alternatives fixed, vary only the decision deadline, and the circuit whose activity predicts, and whose perturbation changes, the response should shift.

That such tests are rare is not an accident: much of the standard network toolkit quietly encodes the hypothesis under dispute. Community detection by modularity maximization defines a good decomposition as one maximizing within- over between-group connections [79], near-decomposability operationalized as an algorithm [80], so using it to ask whether the brain is near-decomposable presupposes the answer. Methods that do not impose assortativity recover core–periphery and disassortative communities that modularity maximization cannot represent at all [81]. Monosynaptic reasoning underestimates influence, which travels multi-step routes [82,83]: the amygdala reaches roughly 90% of prefrontal cortex after one additional connection though directly linked to about 40% [84], and its functional connectivity is better predicted by communicability than by direct pathways [85]. Cortico-centric analysis begs the question, since a claim about the failure of a cortical apex cannot be evaluated on cortex-only data [86,87]. Session-averaged functional connectivity may estimate a quantity corresponding to no configuration the system ever occupied [88], and the same anatomical circuit can express several distinct effective-connectivity patterns depending on its dynamical state, with no change in structural connections at all [89]. Dynamic causal modeling does not forbid cycles, having been introduced to model loops and reciprocal connections but it compares a model space the investigator specifies [90], so cyclic alternatives are examined only when someone decides to include them.

## 6. Concluding remarks

If control is heterarchical, several long-standing puzzles change shape. The question is no longer which structure owns a capacity, but which configuration governs on a particular occasion. Practice can change which circuit controls an action while displaced solutions remain available; task phase can reverse causal direction within a loop; urgency and affordances can recruit different routes through the same anatomy. These are not exceptions to a chain of command. They follow from an architecture in which control relations are assembled from reciprocal pathways whose influence depends on history, context, and dynamical state.

The reframing is not a retreat into holism: regions differ, projections are asymmetric, and some structures are likely necessary; function remains anchored in anatomy. What anatomy underdetermines is which relation operates in the moment. The object of explanation therefore shifts from the region to the temporal evolution of the joint state, replacing “what is the controller?” with “what configuration is the system in, and what put it there?” Dynamical decoupling supplies one axis of the answer: as urgency changes, so does how far history, internal state, and anticipated consequences can shape commitment. Testing the account requires models whose nodes and edges can change with context, rather than models built on fixed graphs [91], and experiments that seek reversals and cycles rather than assume a ranking. Biological control is not a chain of command but a dynamics of mutually constraining mechanisms, each embedded in what it regulates, none sovereign, whose ordering belongs to the occasion rather than the system.

# Box 1. Historical background

For a century, research on the brain basis of emotion argued about which structures matter: Bard placed the hypothalamus at the epicenter [92], Papez proposed a specific but distributed circuit [93], MacLean consolidated a “limbic system” [94]. Cutting across that debate sat an assumption receiving little consideration: whichever structures are involved, the higher ones govern the lower. It survives intact in the canonical modern account of prefrontal cortex inhibiting the amygdala [5]. Its clearest statement is far older, in Hughlings Jackson’s nervous system of evolutionary levels, higher centers bearing down on lower and more primitive ones, with disease releasing the lower from control [95]. That picture is why a hierarchical reading can feel like the null hypothesis when it is a substantive, historically loaded commitment [86].

Warren McCulloch (of McCulloch and Pitts fame) rejected that framework in a way that impacted neuroscience more broadly. He observed that “hierarchy” is a sacerdotal term: on the model of ecclesiastical rank, “the many ends are ordered by the right of each to inhibit all inferiors” [96]. He then gave a precise reason to doubt that nervous systems are so ordered, and it came from where an ordering would have to live. Reflexes are closed circuits rather than arcs; without influence passing between them they “would occur independently of one another,” so it is to those between-circuit influences that “one must look for sub-, super-, or co-ordination of reflexes.” But those same influences can close into a cycle, and a cycle admits no ranking. Experimental work on choice behavior had reported circular preference under constant conditions: an animal prefers A to B, B to C, and yet C to A. If goals lay on a single scale this could not happen, since preferring A to B and B to C would settle that A is preferred to C. So there is no such scale. McCulloch showed that any network producing such behavior must contain a loop running between the competing circuits rather than within any one of them. He called that loop a diadrome and the organization a *heterarchy*.

Two features of the result matter for the present argument. First, it is topological: what makes a global ranking unavailable is a cycle in the graph of between-circuit influence, exactly the property that premise (iii) of the main text asserts of constraint order. Second, the result does not depend on the influences between circuits being inhibitory. Answering questions raised about the original paper, McCulloch showed that connections between circuits that are merely summative, each singly insufficient to fire the next neuron but reinforceable from another circuit, produce the same circularity [97]. What defeats the ranking is the loop, not the sign of the couplings that close it.

# Box 2. Four conceptual anchors of the framework

**Historicity.** What a brain does now depends on what it has done before. Evolution and development establish a repertoire of pathways; learning alters their weighting, coupling, and recruitment. Animals with similar macroanatomy can therefore differ in which circuit governs the same behavior. The return of a behavior need not restore its original control configuration: after looming extinction, defensive responding can return without recovery of the midbrain responses that first controlled it. Present relations therefore embody histories unfolding over several timescales. Section 3 provides examples.

**Evolving network of reciprocal control.** Control belongs to relations among mechanisms, not to structures considered in isolation. A region may set the terms for one process while being constrained in another, and these roles can coexist. Reciprocal wiring alone does not establish heterarchy. What matters is *reciprocal constraint*: structures shape what one another can learn, express, or transmit. When these relations form cycles, genuine local asymmetries cannot be combined into a stable ranking. Section 2 develops this composition argument.

**Context sensitivity.** Context does more than adjust the strength of a fixed relation; it helps determine which relation becomes dominant. It is not merely an external input, but a relation among the situation, the organism's history and state, and its conditions of persistence. What the terrain affords can recruit different escape routes; the time available can favor circuits with different integration demands. The direction of dominant influence may therefore reverse across contexts rather than merely change in magnitude. Section 3 provides examples.

**The spectrum of control.** Configurations vary in how far a response is dynamically decoupled from immediately present conditions. Low urgency gives memory, motivation, and anticipated consequences more time to shape commitment; high urgency favors circuits able to commit through fewer integration steps. These are not two kinds of control or two systems arranged hierarchically. They are positions on a continuum afforded by the same macroarchitecture, with intermediate configurations as the ordinary case. Urgency, understood relative to the organism's viability norms, biases where an occasion falls on the spectrum. Section 4 develops this proposal.

These conceptual anchors are interrelated. History shapes the available relations; context biases which become dominant; reciprocal architecture permits reconfiguration without a standing controller; and dynamical decoupling characterizes the resulting configuration. A hierarchy, when locally useful, describes a time-bounded arrangement, not the brain's enduring organization.

## Outstanding questions

- Can directed influence be estimated separately by context, urgency condition, and learning phase, and tested explicitly for reversal of dominance and for cycles, rather than fit to a model space that assumes acyclicity?
- Is experience-dependent reassignment of control across the neuroaxis (as in looming extinction) a general phenomenon or a special case? What learning conditions produce it, and is it ever reversible?
- Do graded, overlapping community methods applied to data spanning cortex, subcortex, midbrain, and hindbrain recover the non-hub-confined overlap the composition argument predicts?
- Can the "degree of decoupling" be operationalized and measured for an engaged circuit, turning the spectrum of control from a conceptual proposal into a measurable quantity?
- Do other "named hierarchies," the HPA axis, the circadian system, hypothalamic feeding circuits, the motor hierarchy, fail to compose when tested by the same criteria?
- How should predictive-coding and hierarchical active-inference accounts be related to this proposal: does the composition argument apply to their hierarchies, or do they escape it?
- What computational models generate trajectory-level control without presupposing a controller, and what would distinguish them empirically from models that posit one?

# Glossary

**Heterarchy.** An organization in which directional relations of influence exist locally but do not compose into a single ranking of the structures involved, because the relations contain cycles.

**Control hierarchy.** A ranking by command of the structures that support a behavioral capacity, stable across the occasions on which that capacity is exercised; the target of the argument. Global orderings elsewhere in the text are global in this sense, not over the whole brain.

**Second-order constraint.** A mechanism that sets the terms on which some other process runs, rather than carrying out a process itself. Its order is relational (defined with respect to a process), so it does not by itself rank structures.

**Dynamical decoupling.** The degree to which a regulating mechanism has dynamics of its own, insulated from the instantaneous state of what it regulates; the mark distinguishing control from dynamic stability.

**Dynamic stability.** Robustness achieved by feedback internal to a process, with no dedicated, decoupled controller (e.g., an idealized reflex).

**Diadrome.** McCulloch's term for a cycle in the influences running between circuits, the structural correlate of intransitive (heterarchical) value.

**Biological control.** A mechanism determines when and how other mechanisms operate, without carrying out their processes itself, in the service of the organism's conditions of persistence. The sense of control used throughout.

**Cognitive control.** The psychological construct of goal maintenance, conflict resolution, and inhibition, conventionally attributed to prefrontal cortex. On the present account not a separate faculty but biological control operating where decoupling is high.

**Functionally integrated systems.** Large-scale cortical-subcortical circuits that jointly bring behavior about; the unit of analysis in the network view of the brain.

**CRR architecture.** The combinatorial, reciprocal, and reentrant macro-connectivity of the mammalian brain.

**Communicability.** A graph measure of influence that sums contributions over all paths, not only direct connections.

**Viability norms.** The organism's own standards of self-maintenance, relative to which urgency and value are defined.